\documentclass[letterpaper, 10 pt, conference]{ieeeconf}  % Comment this line out if you need a4paper

\IEEEoverridecommandlockouts                              % This command is only needed if 
\usepackage{graphics} % for pdf, bitmapped graphics files
\usepackage{amsmath} % assumes amsmath package installed
\usepackage{amssymb}  % assumes amsmath package installed

\usepackage{graphicx}
\usepackage{xcolor}
\newtheorem{remark}{Remark}[section]

\title{\LARGE \bf
Physics--guided neural networks for inversion--based feedforward control applied to hybrid stepper motors*
}

\author{D. Fan$^{1}$, M. Bolderman$^{\dagger,1}$, S. Koekebakker$^{2}$, H. Butler$^{1,3}$, and M. Lazar$^{1}$% <-this % stops a space
\thanks{*This work is supported by the NWO research project PGN Mechatronics, project number 17973.}% <-this % stops a space
\thanks{$^{\dagger}$Corresponding author: {\tt\small m.bolderman@tue.nl} }
\thanks{$^{1}$Control Systems Group, Eindhoven University of Technology, Groene Loper 19, Eindhoven, 5612 AP, The Netherlands}%
\thanks{$^{2}$Canon Production Printing, St. Urbanusweg 43, Venlo, 5900 AE, The Netherlands.}%
\thanks{$^{3}$ASML, De Run 6501, Veldhoven, 5504 DR, The Netherlands.}
}

\begin{document}

\maketitle
\thispagestyle{empty}
\pagestyle{empty}

%%%%%%%%%%%%%%%%%%%%%%%%%%%%%%%%%%%%%%%%%%%%%%%%%%%%%%%%%%%%%%%%%%%%%%%%%%%%%%%%
\begin{abstract}
Rotary motors, such as hybrid stepper motors (HSMs), are widely used in industries varying from printing applications to robotics.
The increasing need for productivity and efficiency without increasing the manufacturing costs calls for innovative control design.
Feedforward control is typically used in tracking control problems, where the desired reference is known in advance.
%Feedforward control is often applied to have systems follow a prespecified reference more accurately. 
In most applications, this is the case for HSMs, which need to track a periodic angular velocity and angular position reference. 
Performance achieved by feedforward control is limited by the accuracy of the available model describing the inverse system dynamics.
In this work, we develop a physics--guided neural network (PGNN) feedforward controller for HSMs, which can learn the effect of parasitic forces from data and compensate for it, resulting in improved accuracy.
Indeed, experimental results on an HSM used in printing industry show that the PGNN outperforms conventional benchmarks in terms of the mean--absolute tracking error. 
\end{abstract}

%%%%%%%%%%%%%%%%%%%%%%%%%%%%%%%%%%%%%%%%%%%%%%%%%%%%%%%%%%%%%%%%%%%%%%%%%%%%%%%%
\section{INTRODUCTION}
\label{sec:Introduction}
Hybrid stepper motors (HSM) are widely used in industrial automation, such as pick--and--place robots~\cite{Sharan2013, Talpur2012}, additive manufacturing~\cite{Kamble2018}, professional printing applications~\cite{Hoang2019}, and more, see, e.g.,~\cite{Acarnley2002} for an overview. 
HSMs can be operated in an open--loop configuration using microstepping~\cite{Derammelaere2014}. 
However, the open--loop stepping often induces unwanted vibrations and is highly inefficient as it applies high currents to be robust for worst case loads. %they require torque design for the maximum load. 
Consequently, for high--precision applications, closed--loop control schemes are often applied in the form of field--oriented control (FOC)~\cite{Kim2011, Le2017}, see also~\cite{Bernardi2022} for control strategies of the inner current control loop. 
Since FOC requires measurements of both the currents and the angular position of the HSM, significant research has been done on sensorless FOC which does not require the additional angular position sensor, see, e.g.,~\cite{Hoang2019, Obermeier1997}. 

For motion control systems, reference tracking performance is typically achieved via feedforward control, while feedback control stabilizes the system and rejects disturbances and feedforward imperfections~\cite{Steinbuch2000}. 
For rotary motors however, the feedforward control design is largely neglected or restricted to be linear, see, e.g.,~\cite{Kim2011} which employs a linear velocity--acceleration feedforward, or~\cite{Wu2017} which employs linear feedforward with a disturbance compensator. 
The complete dynamical behaviour of HSMs constitutes more complex phenomena, such as parasitic torques arising from manufacturing tolerances, as well as torque ripples caused by detent torque and back electromotive forces. 
Since performance achieved by feedforward control is limited by the accuracy of the model of the inverse system~\cite{Devasia2002}, designing a feedforward controller from a linear model intrinsically limits performance. 
Iterative learning control~\cite{Bristow2006} provides the potential to improve tracking performance further, but requires multiple repetitions of the same reference.
%linear feedforward controllers are intrinsically limited in the performance. 

Physics--guided neural networks (PGNNs) have potential to improve performance achieved by linear, physics--based, feedforward controllers by accurately identifying the inverse system dynamics from data~\cite{Bolderman2021}. 
PGNNs effectively merge physics--based and NN--based models and thereby result in nonlinear feedforward controllers with improved performance, and the same reliability as physics--based feedforward controllers~\cite{Bolderman2023}. 
This is in contrast to black box NNs, which can fail to learn from presented data. 
The application of a PGNN feedforward controller to a rotary machine however remains unexplored.

Hence, this motivates us to develop PGNN feedforward controllers for improving performance of HSMs. %, which can potentially impact a large market, including the printing industry. 
To this end, we define a PGNN architecture that embeds a simple, physics--based inverse model of the HSM within a black--box NN.
Also, we impose the rotational reproducible behaviour, i.e., the same dynamics is expected for each rotation. 
%In this work, we develop a PGNN feedforward controller for an HSM. 
%We impose the rotational reproducible behaviour in the PGNN feedforward controller, i.e., the same dynamics is expected for each rotation. 
With the PGNN architecture defined, the PGNN training identifies or learns the inverse system dynamics of the HSM from an available input--output data set, i.e., requiring measurements of the angular position. 
%The PGNN learns the inverse system dynamics of the HSM from an available input--output data set, i.e., we require measurements of the rotation. 
Since the PGNN feedforward controller does not require online measurements, it can also be implemented in a sensorless FOC scheme. 
%Afterwards however, the PGNN feedforward controller can be implemented in a sensorless FOC scheme. 
%Moreover, the PGNN training and data generation are discussed, and performance is validated on a real--life HSM used within the printing industry.
The developed PGNN feedforward improves performance by a factor $2$ in terms of the mean--absolute tracking error (MAE) in real--time on an HSM used in printing industry, without requiring additional computational hardware or measurements. 

%The remainder of this paper is organized as follows: Section~\ref{sec:Preliminaries} introduces the first--principle modeling of an HSM and the control setup.
%The problem statement is formulated in Sec.~\ref{sec:ProblemStatement}, followed by the PGNN feedforward controller design in Sec.~\ref{sec:PGNNFeedforward}. 
%The developed methodology is validated on a real--life HSM in Sec.~\ref{sec:ExperimentalValidation}, and the conclusions are presented in Sec.~\ref{sec:Conclusions}.

%%%%%%%%%%%%%%%%%%%%%%%%%%%%%%%%%%%%%%%%%%%%%%%%%%%%%%%%%%%%%%%%%%%%%%%%%%%%%%%%
\section{PRELIMINARIES}
\label{sec:Preliminaries}

\subsection{First--principle modeling of an HSM}
\begin{figure*}
    \centering
    \includegraphics[width=0.9\linewidth]{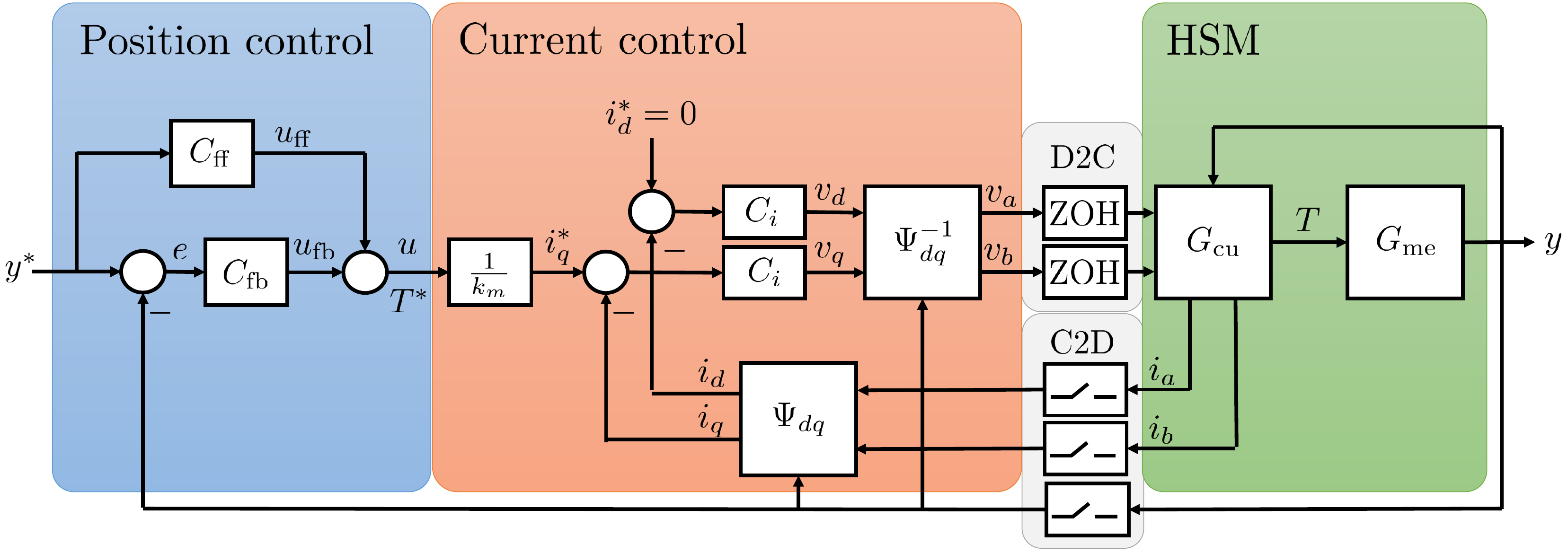}
    \caption{FOC architecture including the HSM, the current control with $dq$--transform and the position feedback--feedforward control setup.}
    \label{fig:ControlScheme}
\end{figure*}

Fig.~\ref{fig:ControlScheme} shows a schematic overview of the FOC structure with $dq$--transformation of the hybrid stepper motor, see, e.g.,~\cite{Rourke2019}.
Note that, both the position and the current controllers are implemented in discrete--time, which are indicated by the sampler and ZOH blocks. 
The HSM is subdivided in a mechanical and an electromagnetic part. 
The mechanical dynamics, indicated with $G_{\textup{me}}$ is modeled using Newton--Euler relations, such that
\begin{equation}
\label{eq:HSMMechanicalDynamics}
    J\frac{d^2}{dt^2} y(t) = F(t) - f_v \frac{d}{dt} y(t),
\end{equation}
where $y(t)$ is the position output at time $t \in \mathbb{R}_{>0}$, $J \in \mathbb{R}_{>0}$ the mass moment of inertia, $f_v \in \mathbb{R}_{>0}$ the viscous friction coefficient, and $F(t)$ the driving torque. 
The driving torque is modeled as~\cite{Henke2013}
\begin{equation}
\label{eq:HSMDrivingTorque}
    F(t) = k_m \left( -i_a(t) \sin \big(N y(t) \big) + i_b(t) \cos \big( N y(t) \big) \right),
\end{equation}
where $k_m \in \mathbb{R}_{>0}$ is the motor constant, $N \in \mathbb{Z}_{>0}$ the number of rotor teeth, and $i_a(t)$ and $i_b(t)$ the current through coils $a$ and $b$, respectively. 

The electromagnetic dynamics is modeled as
\begin{align}
\begin{split}
\label{eq:HSMElectromagnetics}
    L \frac{d}{dt} i_a(t) & = v_a(t) - R i_a(t) + k_m \big( \frac{d}{dt} y(t) \big)  \sin \big( N y(t) \big), \\
    L \frac{d}{dt} i_b (t) & = v_b(t) - R i_b(t) - k_m \big( \frac{d}{dt} y(t)\big)   \cos \big( N y(t) \big),
\end{split}
\end{align}
where $L \in \mathbb{R}_{>0}$ is the inductance, $R \in \mathbb{R}_{>0}$ the resistance, and $v_a(t)$ and $v_b(t)$ the terminal voltages of coil $a$ and $b$, respectively. 
The latter terms in~\eqref{eq:HSMElectromagnetics} are the self--induced voltage, also known as the back electromotive force. 

Since the HSM has two inputs, i.e., the voltages $v_a$ and $v_b$, and only a single output $y$, often the $dq$--transformation~\cite{Rourke2019} $\Psi_{dq}$ is employed
\begin{align}
\begin{split}
\label{eq:HSMDQTransformation}
    \begin{bmatrix} i_d(t) \\ i_q(t) \end{bmatrix} &= \Psi_{dq} \big( y(t) \big) \begin{bmatrix} i_a(t) \\ i_b (t) \end{bmatrix} \\
    :&= \begin{bmatrix} \cos\big( Ny(t) \big) & \sin \big( N y(t) \big) \\ - \sin \big( N y(t) \big) & \cos \big( N y(t) \big) \end{bmatrix} \begin{bmatrix} i_a(t) \\ i_b (t) \end{bmatrix}. 
\end{split}
\end{align}
As a result, we observe that the driving torque in~\eqref{eq:HSMDrivingTorque} simplifies to $T(t) = k_m i_q(t)$, and the mechanical dynamics into
\begin{equation}
\label{eq:HSMMechanicalDynamicsDQTransformed}
    J \frac{d^2}{dt^2} y(t) = k_m i_q(t) - f_v \frac{d}{dt} y(t). 
\end{equation}
Note, from~\eqref{eq:HSMMechanicalDynamicsDQTransformed} we observe that the position control only requires $i_q(t)$. 
Finally, energy consumption can be approximated by the squared sum of currents, which, using $dq$--transformation~\eqref{eq:HSMDQTransformation}, yields
\begin{equation}
\label{eq:HSMEnergyConsumption} 
    i_a^2 + i_b^2 = i_d^2 + i_q^2.
\end{equation}
Since $i_d$ does not contribute to the driving torque, we aim to have it equal to zero and thereby minimize the energy consumption. 

\begin{remark}
    It is possible to derive a more complex description of the HSM dynamics, e.g., by including detent torque, reluctance, and other effects. 
    However, the goal of this work is to demonstrate effectiveness of the PGNN framework for feedforward control, which should compensate for unmodeled effects by learning these effects from data.
    %Nevertheless, it is possible to spend more time on developing a more complex model, although these will inherently be limited in accuracy, e.g., real--life systems are always subject to manufacturing tolerances which creates complex to model dynamics. 
\end{remark}

\subsection{Field--oriented control architecture of an HSM}
The inner current control loop aims to have the driving torque $T(t)$ become equal to the input $u(t)$. 
The currents $i_a(t)$ and $i_b(t)$ are controlled using the voltages $v_a(t)$ and $v_b(t)$, such that, in $dq$--coordinates these follow the references $i_d^*(t) = 0$ and $i_q^*(t)  = \frac{1}{k_m} u(t)$. 
In order to achieve this, the inverse $dq$--transformation is applied to the voltages, such that
\begin{align}
\begin{split}
\label{eq:HSMInverseDQTransformation}
    \begin{bmatrix} v_a(t) \\ v_b(t) \end{bmatrix} &= \Psi_{dq}^{-1} \big( y(t) \big) \begin{bmatrix} v_d(t) \\ v_q(t) \end{bmatrix} \\
    & = \begin{bmatrix} \cos \big( N y(t) \big) & - \sin \big( N y(t) \big) \\ \sin \big( N y(t) \big) & \cos \big( N y(t) \big) \end{bmatrix} \begin{bmatrix} v_d(t) \\ v_q(t) \end{bmatrix}.
\end{split}
\end{align}
Substituting the $dq$--transformation~\eqref{eq:HSMDQTransformation} and the inverse $dq$--transformation~\eqref{eq:HSMInverseDQTransformation} in the electromagnetic model~\eqref{eq:HSMElectromagnetics}, gives the electromagnetic model in $dq$--coordinates as
\begin{align}
\begin{split}
\label{eq:HSMElectromagneticsDQTransformation}
    L \frac{d}{dt} i_d (t) & = v_d(t) - R i_d(t) + LN i_q(t) \frac{d}{dt} y(t), \\
    L \frac{d}{dt} i_q (t) & = v_q(t) - R i_q(t) - k_m \frac{d}{dt} y(t) - LN i_d \frac{d}{dt} y(t).
\end{split}
\end{align}
The voltages in $dq$--coordinates are computed using the discrete--time feedback controller $C_i(z)$ as
\begin{align}
\begin{split}
\label{eq:CurrentControllers}
    v_d(k) & = -C_i (z) i_d(k), \\
    v_q(k) & = C_i (z) \big( i_q^*(k) - i_q(k) \big),
\end{split}
\end{align}
where $k \in \mathbb{Z}_{\geq 0}$ indicates the discrete--time instant. 
The inverse $dq$--transformation $\Psi_{dq}^{-1}$ in~\eqref{eq:HSMInverseDQTransformation} and $dq$--transformation in~\eqref{eq:HSMDQTransformation} are evaluated at discrete time indices, i.e., for $t = kT_s$, with $T_s \in \mathbb{R}_{>0}$ the sampling time.   
The feedback controller $C_i(z)$ is a discretized version of the PI--controller
\begin{equation}
\label{eq:CurrentControllerTransferFunction}
    C_i(s) = k_p + \frac{k_i}{s},
\end{equation}
with $k_p \in \mathbb{R}$ and $k_i \in \mathbb{R}$ the proportional and integral gain, respectively. 

\begin{remark}
    The use of the $dq$--transformation can be omitted by directly transforming the control input $u(k)$ into current references $i_a^*(k)$ and $i_b^*(k)$, see~\cite{Kim2011}. 
    When following a constant velocity reference and assuming a constant load (viscous friction), both $i_a^*(k)$ and $i_b^*(k)$ follow a sinusoidal reference, whereas $i_q^*(k)$ remains constant. 
    Correspondingly, the $dq$ current control is expected to work better for reference tracking control.
\end{remark}

The outer angular position control loop consists of a feedback and a feedforward controller, such that
\begin{equation}
\label{eq:PositionControl}
    u(k) = u_{\textup{fb}}(k) + u_{\textup{ff}}(k),
\end{equation}
where $u_{\textup{fb}}(k)$ is the feedback and $u_{\textup{ff}}(k)$ the feedforward input. 
The feedback input is computed as
\begin{equation}
\label{eq:PositionControlFeedback}
    u_{\textup{fb}}(k) = C_{\textup{fb}}(z) \big( y^*(k) - y(k) \big),
\end{equation}
where $C_{\textup{fb}}(z)$ is the transfer function of the discrete--time feedback controller, and $y^*(k)$ the reference.

We develop a data--driven feedforward controller following the same steps as in~\cite{Bolderman2021}, where linear motors were considered. 
First, we have an input--output data set generated on the system, i.e.,
\begin{equation}
\label{eq:DataSet}
    Z^N := \{ u_0, y_0, ..., u_{N-1}, y_{N-1} \},
\end{equation}
where $N \in \mathbb{Z}_{>0}$ are the number of samples, and $u_i$, $y_i$ are $u(i)$, $y(i)$ for the data generating experiment. 
Second, we parametrize the inverse system dynamics according to
\begin{align}
\begin{split}
\label{eq:ModelClass}
    \hat{u}\big( \theta, \phi(k) \big) & := f \big( \theta, \phi(k) \big), \\
    \phi(k) & := [y(k+n_k+1), ..., y(k+n_k-n_a+1), \\
    & \quad \quad \quad u(k-1), ..., u(k-n_b+1)]^T. 
\end{split}
\end{align}
In~\eqref{eq:ModelClass}, $\hat{u}$ is the prediction of the input $u$, $f :\mathbb{R}^{n_{\theta} \times (n_a+n_b)} \rightarrow \mathbb{R}$ is a model of the inverse dynamics, $\theta \in \mathbb{R}^{n_{\theta}}$ are the parameters, and $\phi(k)$ is the regressor with $n_a, n_b \in \mathbb{Z}_{\geq0}$ describing the order of the dynamics and $n_k \in \mathbb{Z}_{\geq 0}$ the number of pure input delays. 
The values for $n_a$, $n_b$, and $n_k$ can be obtained, e.g., by discretizing a first--principle model of the continuous--time dynamics, or by analyzing a frequency response function.
In order to have the model~\eqref{eq:ModelClass} fit the inverse system dynamics, the parameters are chosen according to an identification criterion
\begin{equation}
\label{eq:IdentificationCriterion}
    \hat{\theta} = \textup{arg} \min_{\theta} \frac{1}{N} \sum_{i=0}^{N-1} \big( u_i - \hat{u} ( \theta, \phi_i ) \big). 
\end{equation}
Finally, the feedforward controller is obtained by computing the input that is required to follow the reference, such that
\begin{align}
\begin{split}
\label{eq:FeedforwardController}
    u_{\textup{ff}}(k) & = \hat{u} \big( \hat{\theta}, \phi_{\textup{ff}}(k) \big), \\
    \phi_{\textup{ff}}(k) & := [y^*(k+n_k+1),  ..., y^*(k+n_k-n_a+1), \\
    & \quad \quad \quad u_{\textup{ff}}(k-1), ..., u_{\textup{ff}}(k-n_b+1)]^T. 
\end{split}
\end{align}
In order to implement the feedforward controller~\eqref{eq:FeedforwardController}, we assume that reference values up until time $k+n_k+1$ are known at time $k$.

%%%%%%%%%%%%%%%%%%%%%%%%%%%%%%%%%%%%%%%%%%%%%%%%%%%%%%%%%%%%%%%%%%%%%%%%%%%%%%%%
\section{PROBLEM STATEMENT}
\label{sec:ProblemStatement}
The choice of the model class~$f$ in~\eqref{eq:ModelClass} determines the effects to be identified, and, consequently, compensated for by the feedforward controller~\eqref{eq:FeedforwardController}.
%The choice of model class~$f$ in~\eqref{eq:ModelClass} is crucial for the design of the feedforward controller~\eqref{eq:FeedforwardController} as this step determines the effects that are of relevance. 
For mechatronic systems, it is typically assumed that the current loop operates significantly faster compared to the position loop, such that the feedforward controller can be designed solely for the mechanical part of the dynamics, i.e.,~\eqref{eq:HSMMechanicalDynamics} with $T(k) = u(k)$. 
Consequently, using the physical knowledge, a suitable candidate for the model class is given as
\begin{align}
\begin{split}
\label{eq:FeedforwardPhysicalModel}
    \hat{u} \big( \theta, \phi(k) \big) & = f_{\textup{phy}} \big( \theta_{\textup{phy}}, \phi(k) \big) \\
    & = \theta_{\textup{phy}}^T \begin{bmatrix} \delta^2 y(k) \\ \delta y(k) \end{bmatrix},
\end{split}
\end{align}
where $\delta = \frac{q-q^{-1}}{2T_s}$ with $q$ the forward shift operator, such that $\phi(k) = [y(k+2), ..., y(k-2)]^T$.
Additionally, $\theta_{\textup{phy}}$ are the physical parameters which represent the inertia $J$ and viscous friction coefficient $f_v$.

\begin{remark}
    It is possible to use more accurate discretization schemes to find $n_a$, $n_b$, and $n_k$ in~\eqref{eq:FeedforwardPhysicalModel}. 
    For example, ZOH discretization is exact for linear dynamics if the input is kept constant between two consequtive samples. 
    However, the experimental results in Sec.~\ref{sec:ExperimentalValidation} show that the parasitic effects are dominant over the discretization error made in~\eqref{eq:FeedforwardPhysicalModel}. 
    %However, as will be apparent from the experimental results in Sec.~\ref{sec:ExperimentalValidation}, the parasitic effects present are dominant over the discretization errors made in~\eqref{eq:FeedforwardPhysicalModel}. 
    Additionally, this discretization scheme has the advantage that $n_b = 0$, such that the feedforward controller~\eqref{eq:FeedforwardController} is stable. 
    For $n_b >0$,~\cite{Bolderman2023} presents tools to both validate (after training) and impose (during training) stability of the PGNN feedforward controllers. 
\end{remark}

\begin{figure}
    \centering
    \includegraphics[width=1.0\linewidth]{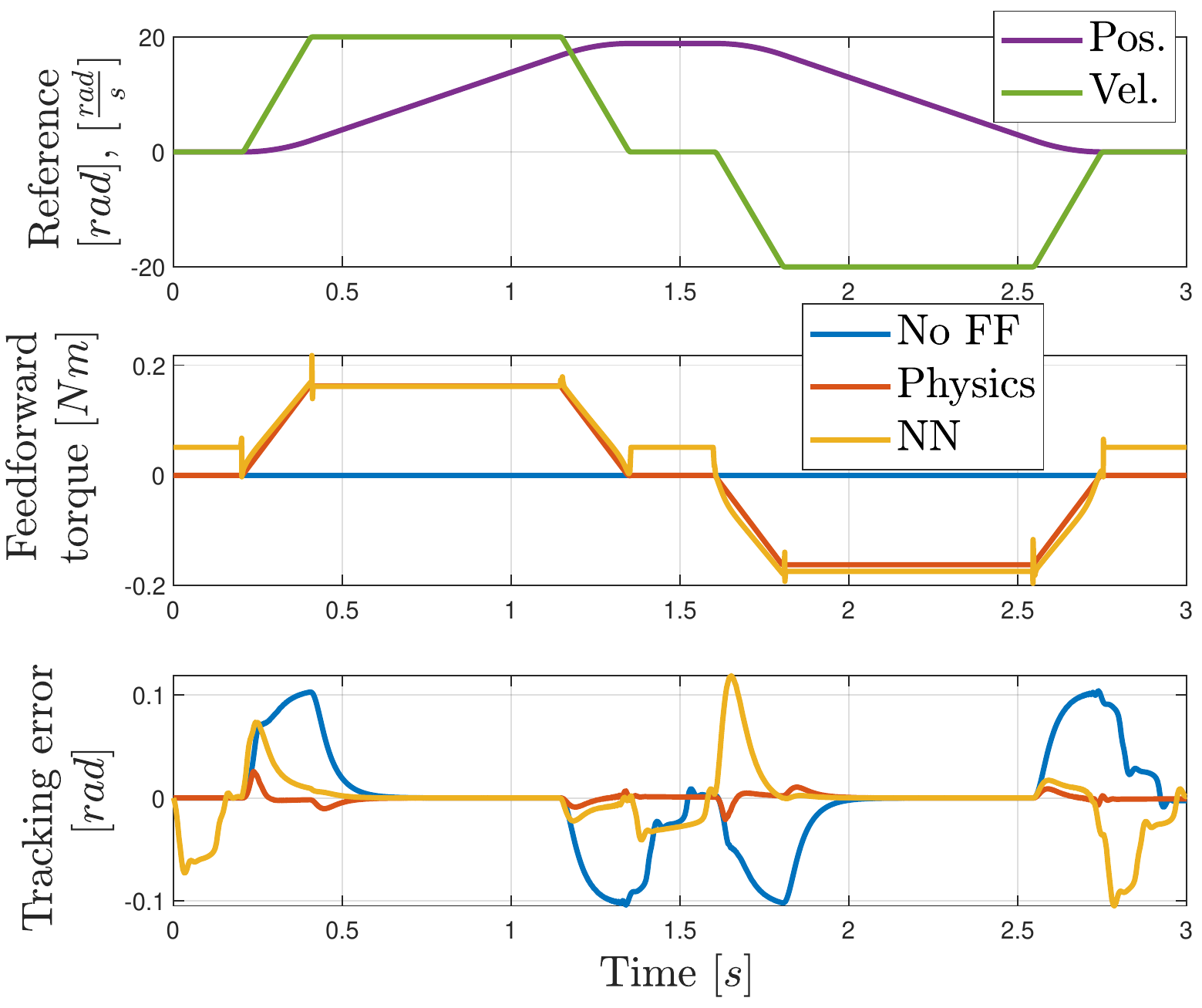}
    \caption{Reference (top window), feedforward signal (middle window), and the resulting tracking error (bottom window) for the feedforward controllers using the physical model~\eqref{eq:FeedforwardPhysicalModel} and the NN~\eqref{eq:FeedforwardNeuralNetwork} on a simulation example.}
    \label{fig:Illustrative_Example}
\end{figure}

The physics--based feedforward controller~\eqref{eq:FeedforwardPhysicalModel} can only identify and compensate for the inertia and viscous friction, while real--life applications comprise of more complex behaviour.
Consequently, it was first proposed in~\cite{Sorensen1999} to employ a black--box NN as a model class~\eqref{eq:ModelClass}, such that
\begin{align}
\begin{split}
\label{eq:FeedforwardNeuralNetwork}
    \hat{u} \big( \theta, \phi(k) \big) & = f_{\textup{NN}} \big( \theta_{\textup{NN}}, \phi(k) \big) \\
    & = W_{L+1} \alpha_{L} \Big( ... \alpha_1 \big( W_1 \phi(k) + B_1 \big) \Big) + B_{L+1},
\end{split}
\end{align}
where $\alpha_l : \mathbb{R}^{n_l} \rightarrow \mathbb{R}^{n_{l}}$ denotes the aggregation of activation functions with $n_l \in \mathbb{Z}_{>0}$ the number of neurons in layer $l \in \{0, ..., L \}$, and $L \in \mathbb{Z}_{>0}$ the number of hidden layers. The parameters $\theta_{\textup{NN}} := [\textup{col} (W_1)^T , B_q^T, ..., \textup{col}(W_{L+1})^T, B_{L+1}^T]^T$ are the concatenation of all weights $W_l \in \mathbb{R}^{n_l \times n_{l-1}}$ and biases $B_l \in \mathbb{R}^{n_l}$, where $\textup{col}(W_l)$ stacks the columns of $W_l$.
Although the NN~\eqref{eq:FeedforwardNeuralNetwork} has the potential to approximate the inverse dynamics up to any accuracy, it lacks the robustness of the physics--based model~\eqref{eq:FeedforwardPhysicalModel}. 
For example, the NN easily fails to learn and generalize from the presented data~\cite{Haley1992}.
%For example, the NN easily fails to learn and generalize from the presented data as a result of the non--convex training to get stuck in a local minimum, as well as the poor extrapolation to non--training data~\cite{Haley1992}.

To illustrate this, we make use of a closed--loop simulation model of an HSM with some parasitic friction forces, see~\cite{Fan2022} for details on the parameters, feedback controllers and friction model.
The simulation closely resembles the real--life setup discussed in Sec.~\ref{sec:ExperimentalValidation}, and follows the same data generation experiment.
We employ feedforward controllers~\eqref{eq:FeedforwardController} based on the physical model~\eqref{eq:FeedforwardPhysicalModel} and the NN model~\eqref{eq:FeedforwardNeuralNetwork} with a single hidden layer $L=1$ with $n_1 = 16$ neurons. 
Fig.~\ref{fig:Illustrative_Example} shows the feedforward signal and resulting tracking error resulting from both feedforward controllers on the HSM simulation. 
Even though the physical model~\eqref{eq:FeedforwardPhysicalModel} significantly improves performance with respect to the situation where no feedforward is applied, there remain some errors that are caused by the inability of the physical model to capture the complete dynamics.
The NN~\eqref{eq:FeedforwardNeuralNetwork} on the other hand, has the capability to learn more complex dynamics.
However, the NN fails to learn and generalize from the presented data which is observed by, e.g., the offset during standstill, and the spikes at the start of the acceleration.
This results in poor tracking performance when the NN--based feedforward controller is applied.
This issue might be reduced by using a different training data set, or by adjusting the NN dimensions and regularization parameters. 
However, the example showcases the sensitivity of the NN.

Consequently, the goal of this work is to effectively embed the known physics--based feedforward controller within a NN--based feedforward controller, termed PGNN, to improve the tracking performance of HSMs.  
%Consequently, the goal of this work is to effectively embed the physics with a NN to achieve improved tracking performance on a real--life HSM.
To this end, we will use a two--step sequential procedure: first, we identify the parameters of a physics--based feedforward controller as in~\eqref{eq:FeedforwardPhysicalModel}. 
Second, we train a NN model~\eqref{eq:FeedforwardNeuralNetwork} on the residuals of the identified physics--based model. 
Then, the physics--based model and the NN model are combined in a single PGNN feedforward controller.

\begin{figure}
\centering
\includegraphics[width=0.96\linewidth]{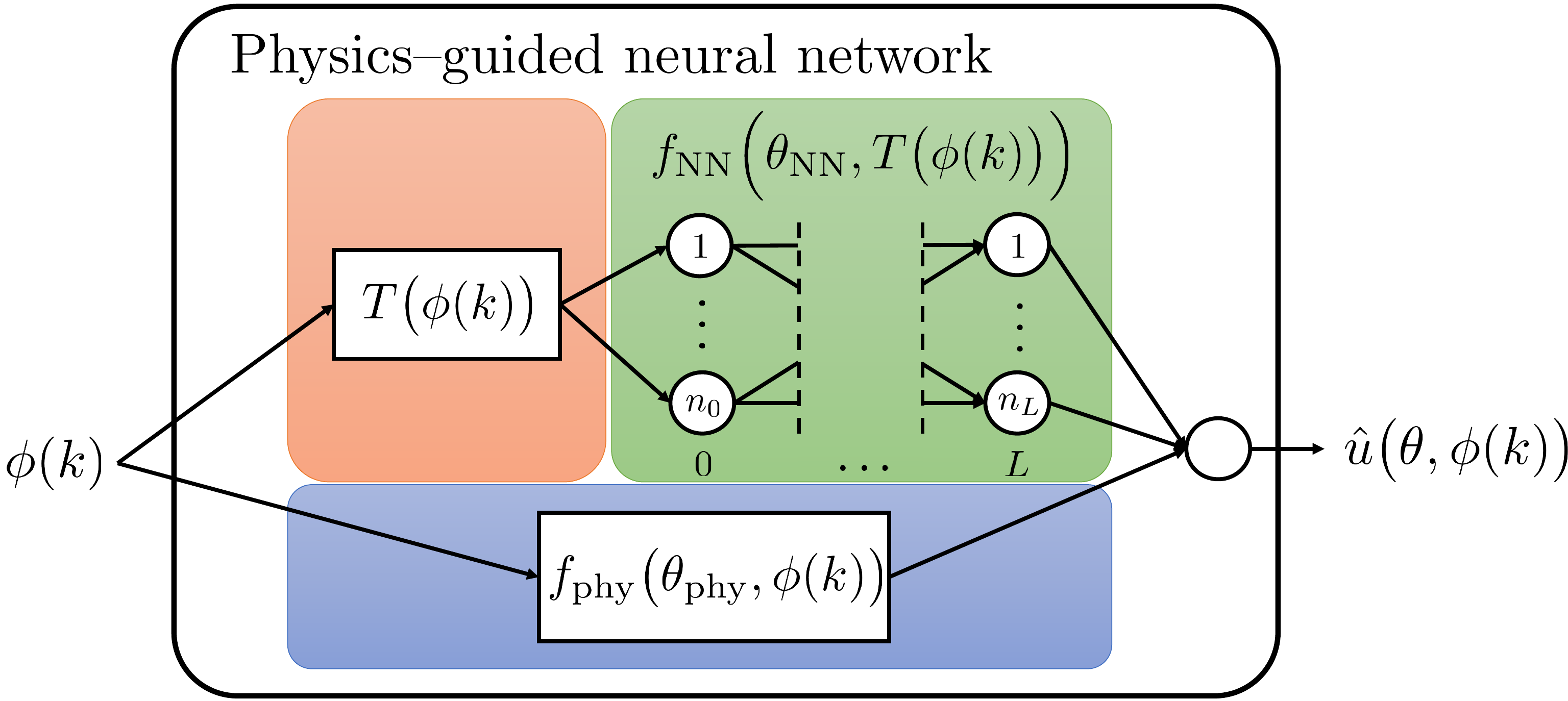}
\caption{Schematic overview of the physics--guided neural network.}
\label{fig:PGNN}
\end{figure}

%%%%%%%%%%%%%%%%%%%%%%%%%%%%%%%%%%%%%%%%%%%%%%%%%%%%%%%%%%%%%%%%%%%%%%%%%%%%%%%%
\section{FEEDFORWARD CONTROL OF HSMS USING PHYSICS--GUIDED NEURAL NETWORKS}
\label{sec:PGNNFeedforward}
With the aim to obtain a feedforward controller with the same reliability as the physics--based model~\eqref{eq:FeedforwardPhysicalModel} and the high accuracy of the NN model~\eqref{eq:FeedforwardNeuralNetwork}, the PGNN model was first proposed in~\cite{Bolderman2021}, see Fig.~\ref{fig:PGNN}. 
The PGNN predicts the input according to
\begin{align}
\begin{split}
\label{eq:FeedforwardPGNN}
    \hat{u} \big( \theta, \phi(k) \big) = f_{\textup{phy}} \big( \theta_{\textup{phy}}, \phi(k) \big) + f_{\textup{NN}} \big( \theta_{\textup{NN}}, T \big( \phi(k) \big) \big), 
\end{split}
\end{align}
where $\theta := [\theta_{\textup{NN}}^T, \theta_{\textup{phy}}]^T$ are the PGNN parameters, and $T : \mathbb{R}^{n_a+n_b} \rightarrow \mathbb{R}^{n_0}$ is an input transformation, with $n_0 \in \mathbb{Z}_{>0}$ the number of NN inputs. 

%The parallel structure of the PGNN~\eqref{eq:FeedforwardPGNN} creates an overparametrization when simultaneously training all parameters, which results in a parameter drift during training. 
%In order to prevent this parameter drift, we employ a two--step identification procedure.
To train the PGNN, we employ the following two--step sequential procedure.
First, the physical parameters $\hat{\theta}_{\textup{phy}}$ are identified according to identification criterion~\eqref{eq:IdentificationCriterion} with physics--based model~\eqref{eq:FeedforwardPhysicalModel}. 
Afterwards, the NN parameters $\hat{\theta}_{\textup{NN}}$ are identified on the residual of the identified physics--based model, such that
\begin{align}
\begin{split}
\label{eq:IdentificationCriterionSequential}
    \hat{\theta}_{\textup{NN}} =& \textup{arg} \min_{\theta_{\textup{NN}}} \frac{1}{N} \sum_{i=0}^{N-1} \big( u_i - \hat{u} ( [\theta_{\textup{NN}}^T, {\hat{\theta}_{\textup{phy}}}^T]^T, \phi_i ) \big) \\
    & \quad \quad \quad \quad \quad + \left\| \Lambda_{\textup{NN}} \theta_{\textup{NN}} \right\|_2^2,
\end{split}
\end{align}
where $\Lambda_{\textup{NN}}$ is a regularization matrix. 
Note that, first identifying $\hat{\theta}_{\textup{phy}}$ and then identifying $\hat{\theta}_{\textup{NN}}$ with $\hat{\theta}_{\textup{phy}}$ fixed can yield a suboptimal solution. This is prevented by identifying $\hat{\theta}_{\textup{phy}}$ and $\hat{\theta}_{\textup{NN}}$ simultaneously as in, e.g.,~\cite{Bolderman2023}.

\begin{figure}
\centering
\includegraphics[width=0.95\linewidth]{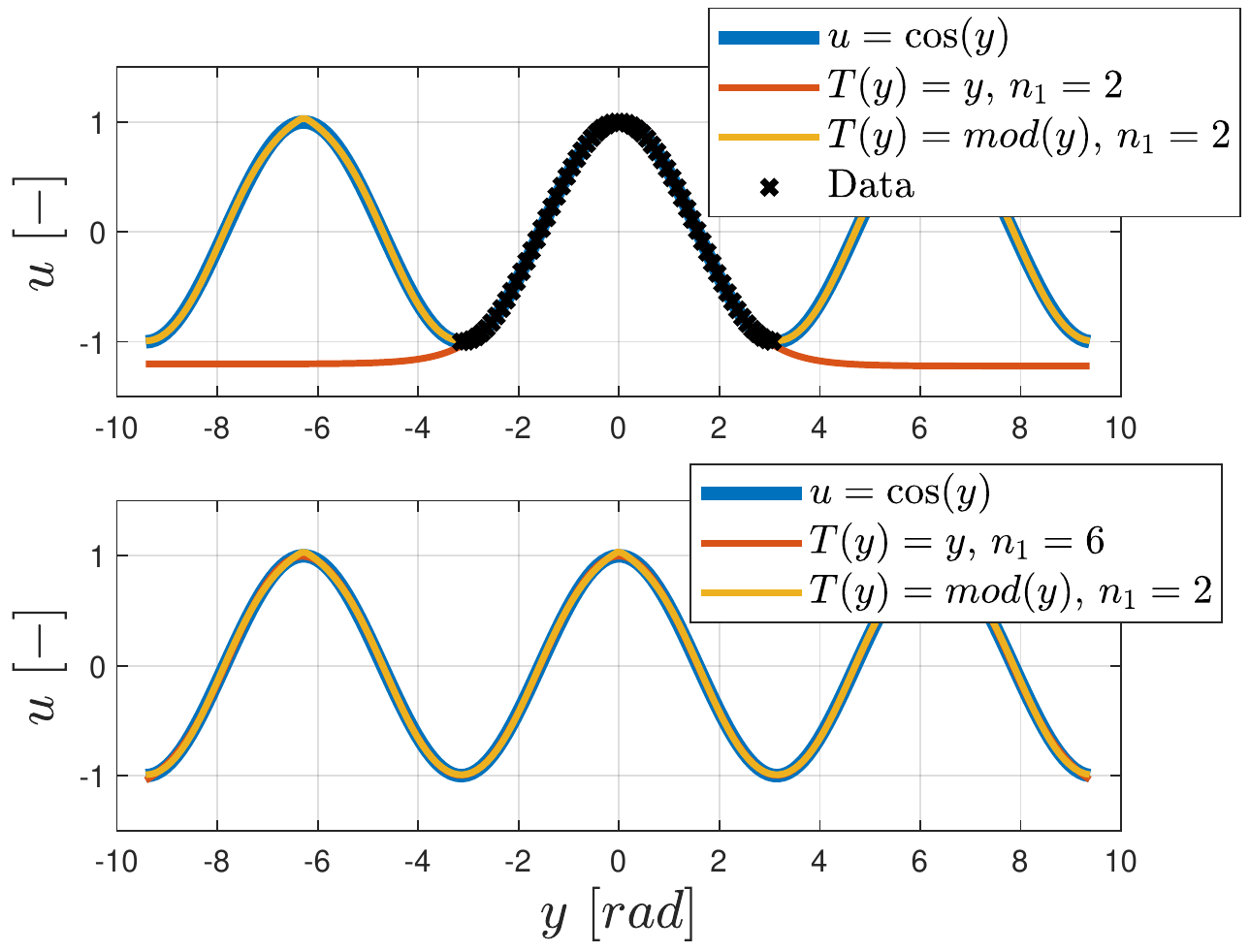}
\caption{Example for imposing physical knowledge via $T(\cdot)$, i.e., improved extrapolation capabilities when training a NN with $T(y) = \textup{mod}(y)$ compared to $T(y) = y$ on a limited data set (top window), and the reduction of the required amount of neurons $n_1$ to achieve an approximation of similar quality (bottom  window).}
\label{fig:Extrapolation_Example}
\end{figure}

It is expected that, for each rotation, the HSM exhibits the same dynamical behaviour. 
It is crucial that the PGNN~\eqref{eq:FeedforwardPGNN} incorporates this rotational reproducibility, since it is otherwise difficult to generate a training data set which describes all relevant rotations $y^*(k)$, e.g., when the HSM rotates in one direction.
%Since HSMs can be operated in one rotational direction, i.e., $y^*(k)$ only increases, it is crucial that the PGNN~\eqref{eq:FeedforwardPGNN} incorporates this rotational reproducibility. 
%Note that it is otherwise not possible to generate a training data set for the PGNN which incorporates all relevant rotations $y^*(k)$. 
Therefore, we aim to identify a PGNN model~\eqref{eq:FeedforwardPGNN} which satisfies
\begin{equation}
\label{eq:ReproducibleRotational}
    \hat{u} \big( \theta, \phi(k) \big) = \hat{u} \left( \theta, \phi(k) + \begin{bmatrix} 1^{(n_a+1) \times 1 } \\ 0^{(n_b-1) \times 1} \end{bmatrix} n 2 \pi \right), \; n \in \mathbb{Z}. 
\end{equation}
In order to impose the rotational reproducible behaviour, i.e., to make the PGNN~\eqref{eq:FeedforwardPGNN} comply with~\eqref{eq:ReproducibleRotational}, we consider a specific design of the physics--guided input transform $T(\cdot)$. 
To do so, we restate that the system order was approximated as $n_k = 1$, $n_a = 4$, and $n_b = 0$ from the physical model~\eqref{eq:FeedforwardPhysicalModel}, such that we consider the following physics--guided input transform
\begin{align}
\begin{split}
\label{eq:PhysicsGuidedInputTransform}
    T \left( \begin{bmatrix} y(k+2) \\ \vdots \\ y(k-2) \end{bmatrix} \right) = \begin{bmatrix} \delta^2 y(k) \\ \delta y(k) \\ \textup{mod} \big( y(k), 2 \pi \big) \end{bmatrix},
\end{split}
\end{align}
where $\textup{mod}\big( y(k), 2\pi \big)$ is the remainder after division of $y(k)$ with $2 \pi$.
Note that,~\eqref{eq:PhysicsGuidedInputTransform} is adopted rather than wrapping all $y$ into the domain $[0, 2\pi)$, since $\delta \textup{mod} \big( y(k) \big) \neq \delta y(k)$ for all $k$.
%Note that it is not suitable to use $\textup{mod} \big( y(k) \big)$ for all time entries, since $\delta \textup{mod} \big( y(k) \big) \neq \delta y(k)$ for all $k$.
Therefore, $T(\cdot)$ includes discrete--time approximations of derivatives of the output $y(k)$, which can also improve the training convergence, e.g., when high sampling rates with respect to the velocity are taken, such that $y(k) \approx y(k-1)$. 

\begin{remark}
    %The PGNN~\eqref{eq:FeedforwardPGNN} with physical model~\eqref{eq:FeedforwardPhysicalModel} and NN using transform~\eqref{eq:PhysicsGuidedInputTransform} satisfies~\eqref{eq:ReproducibleRotational}.
    The physical model~\eqref{eq:FeedforwardPhysicalModel} only inputs discrete--time angular velocity and acceleration, such that it is reproducible for any offset $y(k) + \Delta$. 
    Then, combined with the NN using transform~\eqref{eq:PhysicsGuidedInputTransform}, the PGNN~\eqref{eq:FeedforwardPGNN} satisfies~\eqref{eq:ReproducibleRotational}.
\end{remark}

As an example of the physics--guided input transform $T (\cdot)$, consider the situation in which a NN is used to learn
\begin{equation}
\label{eq:Cosine}
    u(k) = \cos \big( y(k) \big),
\end{equation}
with data generated from one period. 
The top window in Fig.~\ref{fig:Extrapolation_Example} shows that $n_1 = 2$ hidden layer neurons (with $\tanh$ activation) give a reasonably accurate identification. 
The lack of data however, causes the NN trained with $T \big( y(k) \big) = y(k)$ to extrapolate poorly, in contrast to the NN trained with $T \big( y(k) \big) = \textup{mod} \big( y(k) \big)$. 
On the other hand, when the full range of interest is covered with data, the NN with $T \big( y(k) \big) = \textup{mod} \big( y(k) \big)$ requires significantly less neurons compared to the NN with $T \big( y(k) \big) = y(k)$ to yield an approximation of similar accuracy, see the bottom window of Fig.~\ref{fig:Extrapolation_Example}.

\begin{figure}
\centering
\includegraphics[width=0.45\linewidth]{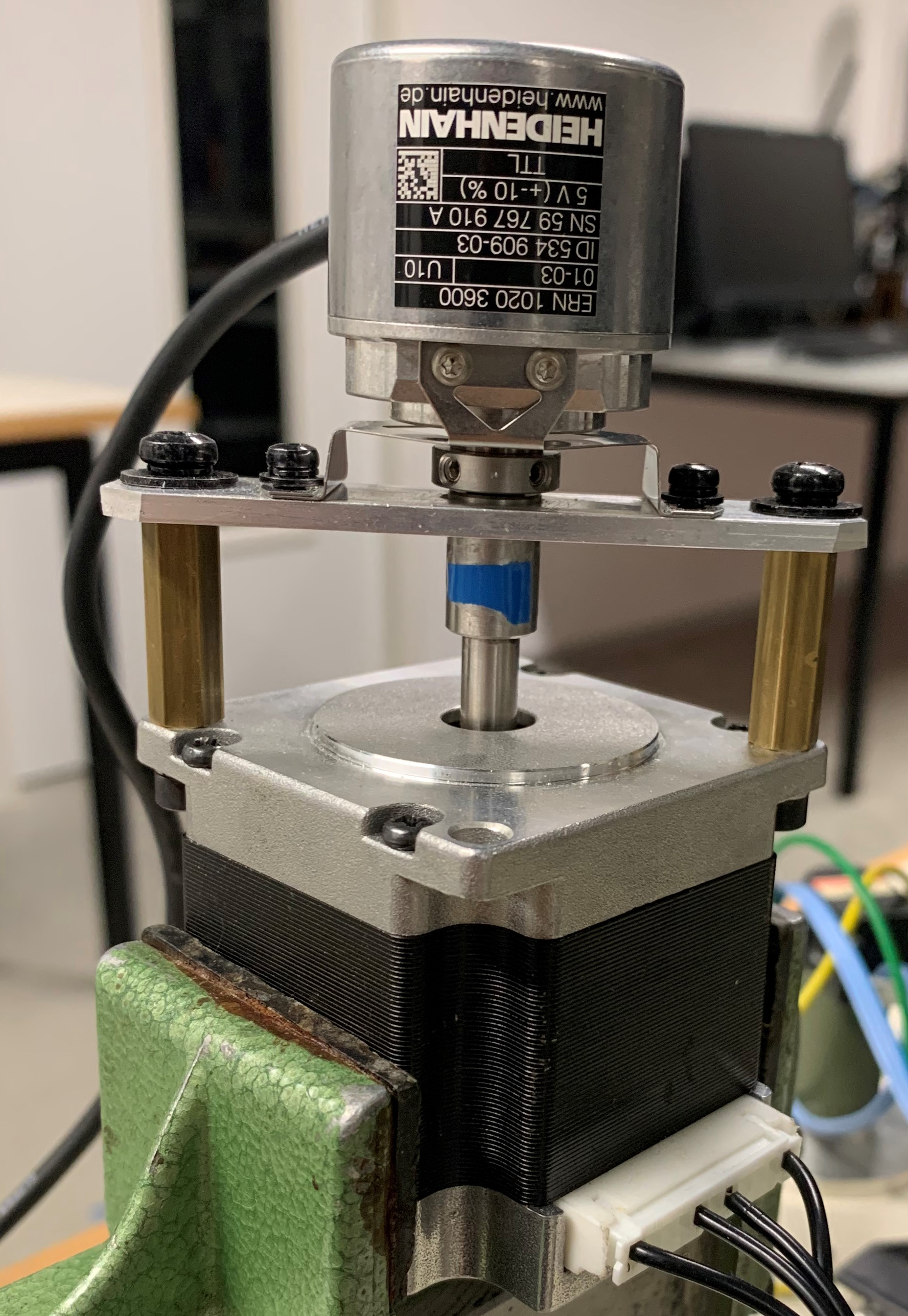}
\caption{HSM FL57STH51--2804A by FULLING MOTOR with encoder.}
\label{fig:HSM}
\end{figure}

%%%%%%%%%%%%%%%%%%%%%%%%%%%%%%%%%%%%%%%%%%%%%%%%%%%%%%%%%%%%%%%%%%%%%%%%%%%%%%%%
\section{EXPERIMENTAL VALIDATION}
\label{sec:ExperimentalValidation}
The PGNN--based feedforward controller~\eqref{eq:FeedforwardPGNN} is validated on a real--life HSM used in printing industry shown in Fig.~\ref{fig:HSM}. 

For simplicity, the current and position controllers are only proportional gains tuned as
\begin{align}
\begin{split}
    \label{eq:ControllersSetup}
    C_i(s) &= 6.6, \quad C_{\textup{fb}} (s) = 5. 
    %C_{\textup{fb}} (s)  &= \frac{6.0\cdot 10^{-3} s^2 + 0.59 s + 7.5 }{1.2 \cdot 10^{-5} s^3 + 7.6 \cdot 10^{-3} s^2 + s}. 
\end{split}
\end{align}

Training data is generated by sampling the input $u(k)$ and output $y(k)$ with a sampling time of $Ts = 10^{-4}$~s while operating the HSM in closed--loop with a third order reference moving back--and--forth between $-3$ to $+3$ rotations with a velocity of $15$ $\frac{\textup{rad}}{\textup{s}}$, acceleration of $80$ $\frac{\textup{rad}}{\textup{s}^2}$, and jerk $1000$ $\frac{\textup{rad}}{\textup{s}^3}$ for a duration of $80$~s. 
The PGNN~\eqref{eq:FeedforwardPGNN} uses the physical model~\eqref{eq:FeedforwardPhysicalModel} and a single hidden layer with $16$ $\tanh$ neurons with physics--guided input transform~\eqref{eq:PhysicsGuidedInputTransform} trained according to~\eqref{eq:IdentificationCriterionSequential} with $\Lambda_{\textup{NN}} = 0$. 
It was observed that adding more neurons or hidden layers did not further improve performance.

\begin{figure}
\centering
\includegraphics[width=1.0\linewidth]{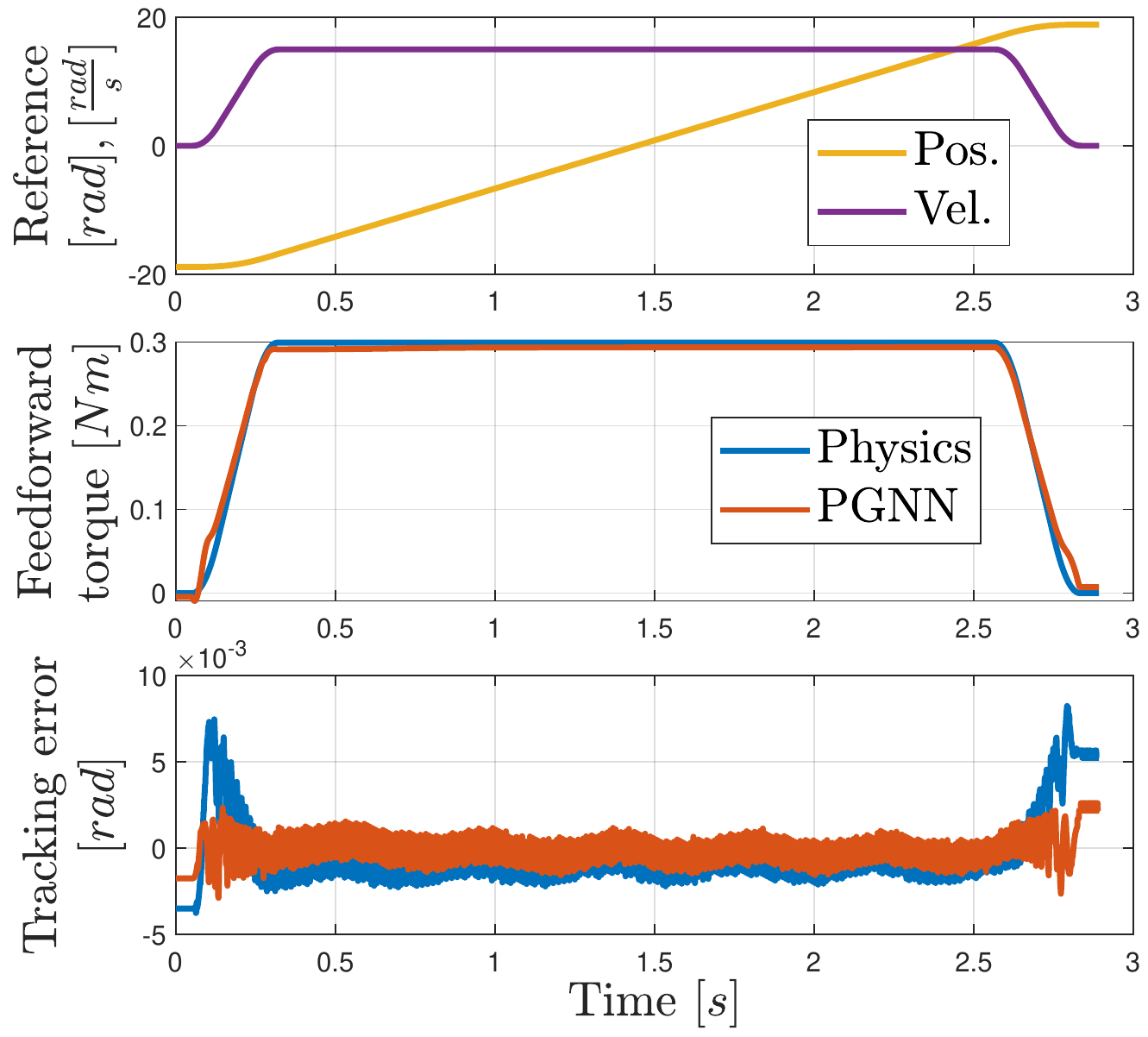}
\caption{Reference (top window), feedforward signal (middle window), and the resulting tracking error (bottom window) for the feedforward controllers using the physical model~\eqref{eq:FeedforwardPhysicalModel} and the PGNN~\eqref{eq:FeedforwardPGNN} for the real--life HSM.}
\label{fig:Experiments_TrackingError}
\end{figure}

Fig.~\ref{fig:Experiments_TrackingError} shows the reference, generated feedforward signals, and the tracking error resulting from the physics--based feedforward and the PGNN. 
The presented forward motion was preceded by a back--and--forward motion of the same reference to remove the transients caused by differences in initial conditions, and thereby facilitate a fair comparison. 
Although the physics--based and the PGNN--based feedforward inputs are largely similar, the small deviations especially during the acceleration part of the reference yield significantly smaller overshoot for the PGNN.

\begin{figure}
\centering
\includegraphics[width=1.0\linewidth]{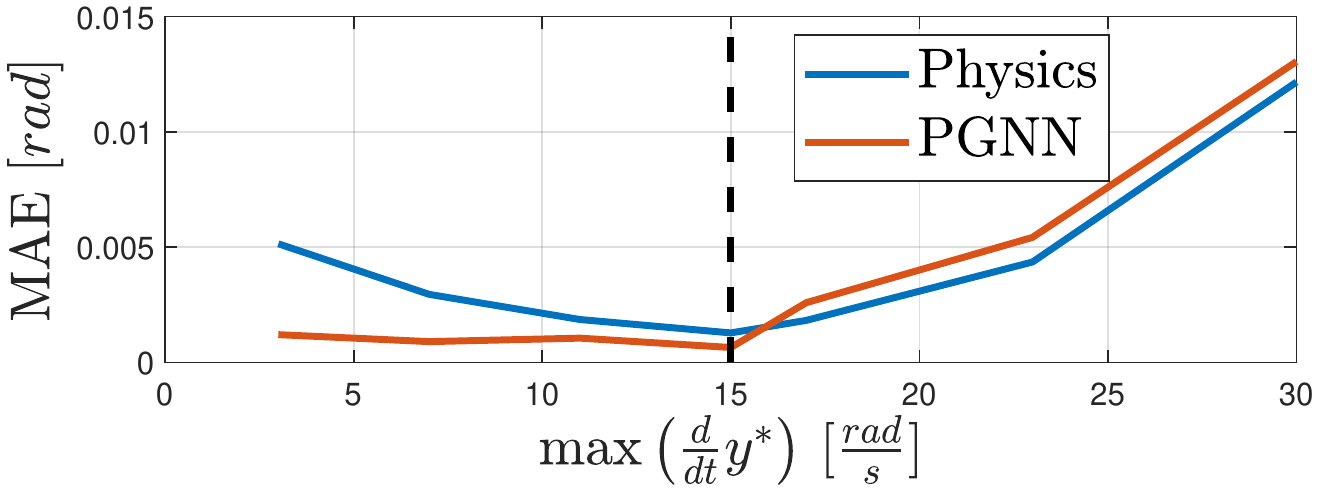}
\caption{MAE of the tracking error using a physics--based feedforward~\eqref{eq:FeedforwardPhysicalModel} and the PGNN~\eqref{eq:FeedforwardPGNN} for references with varying velocities. The black line indicates the maximum velocities attained during training.}
\label{fig:MAE}
\end{figure}

Fig.~\ref{fig:MAE} shows the mean--absolute error (MAE) 
\begin{equation}
\label{eq:MAE}
    \frac{1}{N_r} \sum_{t=0}^{N_r-1} | y^*(k) - y(k) |,
\end{equation}
for a reference of $N_r \in \mathbb{Z}_{>0}$ samples as in Fig.~\ref{fig:Experiments_TrackingError} with different maximum velocities. 
The PGNN outperforms the physics--based feedforward controllers for all velocities smaller than $15$ $\frac{\textup{rad}}{\textup{s}}$. 
For velocities larger than $15$ $\frac{\textup{rad}}{\textup{s}}$, the physics--based feedforward controller achieves slightly better performance, which is explained by the fact that the training data did not contain information for velocities exceeding $15$ $\frac{\textup{rad}}{\textup{s}}$. 
It is possible to enhance robustness to non--training data via the regularization approach discussed in~\cite{Bolderman2022b}, which penalizes the deviation of the PGNN output with respect to the output of the physical model for non--training data.


%%%%%%%%%%%%%%%%%%%%%%%%%%%%%%%%%%%%%%%%%%%%%%%%%%%%%%%%%%%%%%%%%%%%%%%%%%%%%%%%
\section{CONCLUSIONS}
\label{sec:Conclusions}
A PGNN--based feedforward controller for HSMs was developed and tested in real--time experiments. %is developed. 
The PGNN was designed to physically embed the rotational reproducible behaviour of the HSM, which improved performance with respect to a physics--based approach on a real--life HSM without requiring an increase in costs.
Further research will focus on the feedforward controller design for HSMs as part of a complex industrial printer, as well as reducing the effect of predictable disturbances on the closed--loop system.

\section{ACKNOWLEDGEMENTS}
The authors thank Steven Schalm and Will Hendrix for making the HSM setup operational.

\bibliographystyle{IEEEtran}
\bibliography{IEEEabrv,References}
\end{document}